\documentclass[aps,pre,showpacs,amssymb,a4paper,twocolumn]{revtex4-1}%
\usepackage{amsmath}%
\usepackage{amsfonts}%
\usepackage{amssymb}%
\usepackage{graphicx}
\usepackage[english]{babel}
\usepackage{color}

\begin{document}

\title{Devising a protocol-related statistical mechanics framework for granular materials} 
\author{Fabien Paillusson} 
\affiliation{Departament de fisica fonamental, Universitat de Barcelona, 1 Marti i Franques, 08028, Barcelona, Spain}
\date{\today}

\begin{abstract}
Devising a statistical mechanics framework for jammed granular materials is a challenging task as those systems do not share some important properties required to characterize them with statistical thermodynamics tools. In a recent paper [Asenjo et al. PRL, 2014], a new definition of a granular entropy, that puts the protocol used to generate the packings at its roots, has been proposed. Following up these results, it is shown that the protocol used in [Asenjo et al. PRL, 2014] can be recast as a canonical ensemble with a particular value of the temperature. Signature of gaussianity for large system sizes strongly suggests an asymptotic equivalence with a corresponding microcanonical ensemble where jammed states with certain basin volumes are sampled uniformly. We argue that this microcanonical ensemble is not Edwards' microcanonical ensemble and generalize this argument to other protocols.
  \end{abstract}

\maketitle

\section{Introduction}
Granular materials are large assemblies of particles which are athermal and dissipative and as such their statistical properties can hardly be characterized by using the standard tool box of statistical thermodynamics. It has then been considered a challenge in the community to devise a novel statistical framework for these systems that would have both explanatory and predictive powers \cite{Blumenfeld09, Zamponi09, Henkes09, PicaCiamarra12}. 
One of the most appealing routes towards a statistical framework to characterize jammed states is the one proposed by Sam Edwards more than two decades ago which hypothesizes that if one has a protocol to generate (granular) packings of $N$ particles at some fixed volume $V$ or packing fraction $\phi$, then each jammed packing will be as likely as any other to appear with probability $\Omega_{jammed}^{-1}(N,\phi)$ --- where $\Omega_{jammed}(N,\phi)$ is the total number of mechanically stable structures that can be made out of these $N$ particles confined in the prescribed volume $V$ --- \citep{Edwards89, Edwards90}. Although such a claim does not bear any micro-mechanical justification yet, there is still a possibility that, akin to throwing a six facet dice (system for which we do not have any micro-mechanical justification either), it so happens that most protocols could well be described by a probability measure which assumes a uniform sampling of the states in virtue {\it e.g.} of Pascal's principle of indifference. Still, Edwards' proposal contrasts deeply with studies like that of Jiao et al. \cite{Torquato2010, Torquato2011} which showed that one could tailor many packing-generating protocols that lead to any possible packing fraction corresponding to various degrees of structural randomness for the obtained jammed states (hence ruling out {\it e.g.} packings with the same packing fraction but with a different randomness). Also, recent advances on the behaviour of hard sphere glasses have permitted to probe the glass phase close to the jamming transition. This probing can be done in two different ways: either by focusing on the ``\ typical ''\  (in a statistical thermodynamic sense) states that can exist close to jamming \cite{Charbonneau14} or by following initial states from a dilute phase \cite{Krzakala10} up to jamming \cite{Rainone14}. Interestingly, these two different ways of choosing glassy states yield a different physics which strengthen the idea of what seems to be a protocol dependent measure. It is therefore important to devise tests to assess whether Edwards' conjecture is rather the rule or the exception to the rule. To this aim,  different strategies have been used. On the one hand, instead of testing Edwards' {\it microcanonical} hypothesis, it has been more practical to test a {\it canonical} extension of it which ought to apply to packing-generating protocols that do not fix exactly the volume occupied by the jammed states. In this case however the various theoretical \cite{Barrat2000, Dean03, Blumenfeld03, Eastham06, Bowles11,Blumenfeld12}, numerical \cite{Aste08, PicaCiamarra06, McNamara09, Paillusson12} and experimental \cite{Makse02, McNamara09, Lechenault10, Makse12, Daniels13} studies done on the subject were not, overall, very conclusive. That is because different protocols and assumptions were used to apply Edwards' canonical ensemble to practical cases. On another hand, it has been proposed for a long time to use soft materials as models for granular materials \cite{Nagel03, Makse05, Chen10, Wang10, Brujic11, Sollich12, Makse12, Xu05}. In this case, the idea consists in interpreting the discrete set of jammed granular states as being minima of a potential energy surface (also referred to as an {\it energy landscape}; term that we should use interchangeably with potential energy surface throughout the paper) of soft particles \footnote{By doing so however, one in principle generates additional spurious energy minima that do not correspond to any granular jammed state. If one wants to keep a fixed packing volume, it is therefore important to design a potential energy surface that tries to minimize the number of these spurious minima so as to get soft jammed states as close as possible to that of granular jammed states. Otherwise, one can use less carefully chosen soft matter models to mimic closely granular jammed states by allowing their packing volume to vary \cite{Xu05}.}. This allows then both experimental \cite{Gao09, Brujic11} and numerical \cite{Makse12,PRL14,Nagel03, Gao06} realizations of soft material models of granular packings that can be used in principle to test Edwards' original hypothesis. 
\begin{figure}
\includegraphics[width=0.99\columnwidth]{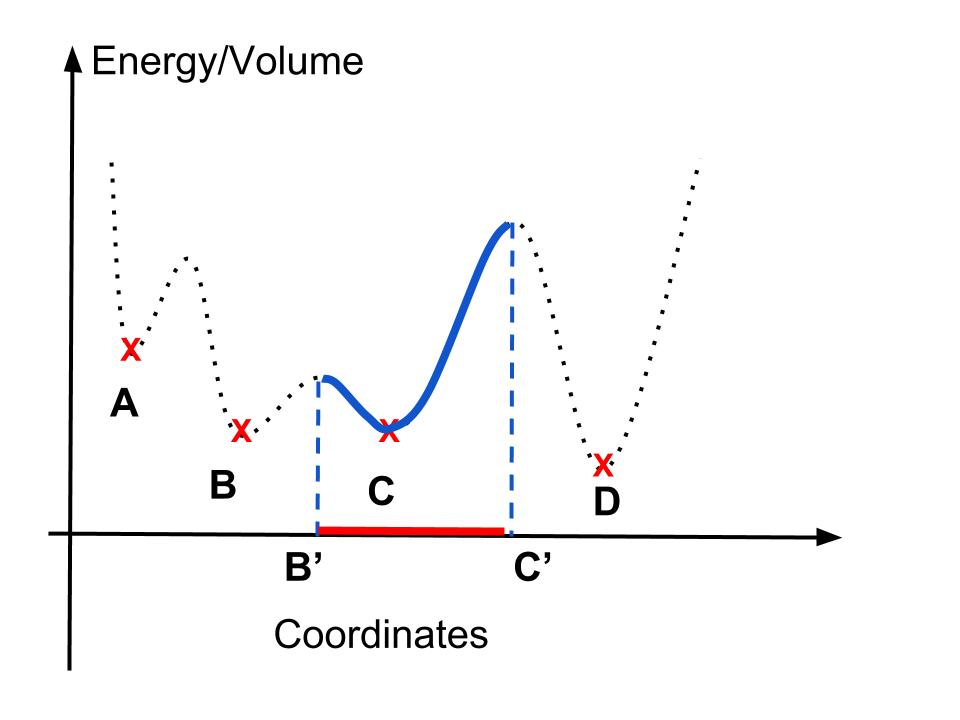}  
\caption{\label{fig1} (Color online) Schematic representation (dotted line) of an energy or volume landscape that can be used to generate four mechanically stable packings (A, B, C, D). For each minimum, it is not possible to decrease further the cost function without first increasing the volume available (volume landscape interpretation like the LS algorithm) or the energy of the system (energy landscape interpretation). The solid blue line represents the basin of attraction of the minimum $C$ while the length of the red segment $B'C'$ represents its size. In a multidimensional space, this size becomes a hyper-volume in the space of coordinates. }
\end{figure}

Now, it appears that even the first set of studies looking at granular jammed states with a non-constrained packing fraction can be interpreted in terms of a landscape that would be a {\it volume} or {\it density} landscape \cite{Ashwin12} where the minima of this landscape would be the set of states for which, under a prescribed dynamical model that prevents particle overlaps ({\it e.g.} the Lubachevsky-Stillinger (LS) algorithm \cite{LS1, LS2}), it is impossible to reach a lower volume configuration before first increasing the volume accessible to the particles \cite{Mueggenburg12}. This allows then a rather unified picture of two of the most ubiquitous ways to generate jammed states expressed in the vocabulary of landscape exploration. Fig.\ref{fig1} gives an illustration of the idea underlying this interpretation of jammed states as being either volume minima in a density landscape of hard particles or energy minima of the potential energy surface of a soft material model of granular matter.

\vspace{2mm}

In this article, we will mostly focus on the energy landscape (potential energy surface) exploration as we seek physical consequences --- regarding {\it e.g.} Edwards' conjecture --- of numerical results obtained in \cite{PRL14}. To this end, we first recall very standard results on the equivalence between the canonical and the microcanonical ensembles in the thermodynamic limit. We then move to jammed systems as studied in \cite{PRL14}. After having recalled the protocol they used and the results they obtained, we will show that there is a strong analogy between an infinite number of energy landscape-based protocols and a --- granular --- canonical ensemble. We will moreover show that, akin to what happens in statistical thermodynamics, this canonical ensemble tends toward a microcanonical measure. We discuss the implications of this equivalence and claim that this microcanonical measure, in general, does not coincide with the Edwards' measure.

\section{Equivalence between the microcanonical and canonical ensembles}
This section is dedicated to recall how does the canonical ensemble become equivalent to a microcanonical ensemble in the thermodynamic limit. Although, the derivation that follows can be found in most text books on statistical thermodynamics ({\it e.g.} in \cite{Huang}), we briefly re-derive it here to introduce notations and concepts that will be useful in the remaining of the manuscript.
We start off by writing the partition function in the canonical ensemble, for $N$ particles in a volume $V$ with inverse temperature $\beta = 1/k_BT$, via a sum over energies $E$ i.e.:
\begin{equation} 
Q(N,V,\beta) = \frac{1}{N!} \int_0^{+\infty}  dE \:\omega(E,V,N)e^{-\beta E} \label{eq0}
\end{equation}where $\omega(E,V,N)dE$ is the total number of states in the energy range $[E, E+dE]$. The idea consists then in defining the microcanonical entropy $S(E,N,V)$ via $\omega(E) \equiv e^{S(E,N,V)/k_B}$ and it comes that the partition function reads:
\begin{equation}
Q(N,V,\beta) =  \int_0^{+\infty}  dE \:e^{-\beta A(E,N,V)} \label{eq1}
\end{equation}where $A(E)= E-TS(E)+ k_B T\ln N!$ is the Helmholtz free energy of the system at energy $E$ and where the $+\ln N!$ term ensures extensivity of $A$. To formally address the extensivity property, we introduce the notion that two functions $f(N,V,\beta)$ and $g(N,V,\beta)$ are equivalent in the thermodynamic limit (and noted $f \sim g$) {\it iff} $\lim_{N,V \rightarrow +\infty} \frac{f}{g} = 1$ at fixed packing fraction $\phi \equiv Nv_0/V \sim \mathcal{O}(1)$ ($v_0^{1/d}$ being a length scale related to the size of the particles; where $d$ stands for the spatial dimension). $A$ being extensive, then --- by definition --- there exists a function $a( e,\phi) \sim \mathcal{O}(1)$ such that $A(E) \sim N a( e,\phi)$ and where $e$ and $a$ are the energy and free energy per particle. In the thermodynamic limit, Eq. \eqref{eq1} can then be recast as:
\begin{equation}
Q(N,V,\beta) \sim  N\int_0^{+\infty}  de \:e^{- N \beta a(e,\phi)}. \label{eq2}
\end{equation}Now, if $a$ has a minimum at say $e^*$, then this integral can be evaluated by using a saddle point approximation. Assuming this extremum lies on the real line, the idea is to expand $a(e) = a(e^*) + \frac{1}{2}a''( e^*)( e- e^*)^2 + \mathcal{O}((e-e^*)^3)$. Replacing it in Eq. \eqref{eq2} up to the second order gives: 
\begin{equation}
Q(N,V,\beta) \sim N e^{-N \beta a(e^*)} \int_0^{+\infty}  de \:e^{-\beta N \frac{a''(e^*)(e-e^*)^2}{2}}  \label{eq3}
\end{equation}From Eq. \eqref{eq3}, we see that all the statistics is now essentially in a gaussian form. From very standard statistical mechanics, we know that $Var(E) \propto C_v$ where $C_v$ is the specific heat capacity at fixed volume. Now, the heat capacity is also an extensive quantity and therefore $Var(E) \sim N$. Now, in the case of Eq. \eqref{eq3}, the statistics is that of $e$ and not $E$. This is no issue as we can use the fact that $Var(e) = Var(E)/N^2 \sim 1/N$. It then implies, in case we had any doubts, that $a''(e^*) \sim \mathcal{O}(1)$. Evaluating the gaussian integral then yields:
\begin{equation}
Q(N,V,\beta) \sim e^{-N \beta a(e^*)}\sqrt{\frac{2 \pi N}{K}}  \label{eq5}
\end{equation}where $K \sim \mathcal{O}(1)$ is a constant. Hence, in the thermodynamic limit, one finds that $A(N,V,\beta) \sim N a(e^*) \equiv N (e^* - Ts(e^*))$ (where $s$ contains here part of the $N!$ introduced in the definition of $A$ in Eq. \eqref{eq1}). Everything is thus as if the energy was fixed at the value $N e^*$, from which we subtract the corresponding (extensive) microcanonical entropy. The role of the thermal bath is encapsulated in the possibly complicated relationship between $e^*$ and $\beta$. Now, since the partition function is the generating functional of all the moments of the canonical distribution, we have therefore retrieved that there is a statistical equivalence between the canonical ensemble and a microcanonical one in the thermodynamic limit {\it i.e.} there exists a one to one correspondence between the physics going on at some $\beta$ value in the canonical ensemble and the physics characterized by the corresponding microcanonical ensemble at $E = Ne^*(\beta)$.

\hspace{2mm}

Another way to see this equivalence is to look at the integrand in Eq. \eqref{eq3} and see that we can build a probability distribution for $e$ from it by simply dividing the gaussian weight by the partition function itself. Since we have calculated it in Eq. \eqref{eq5}, we see that at least up to order 2, we have that:

\begin{equation}
p_{N}(e) \approx \sqrt{\frac{K N}{2\pi}} \exp \left( - \frac{K N (e-e^*)^2}{2} \right) \label{eq7}
\end{equation}One can easily check that $p_{N}$ is essentially a sequence of functions whose limit is the Dirac delta distribution such that:

\begin{equation}
\lim_{N \rightarrow +\infty} p_{N}(e) = \delta(e-e^*) \label{eq8}
\end{equation}which is essentially a microcanonical measure at fixed energy $Ne^*$.

\section{Generating jammed states and the landscape ambiguity}
\begin{figure}
(a)\includegraphics[width=0.4\columnwidth]{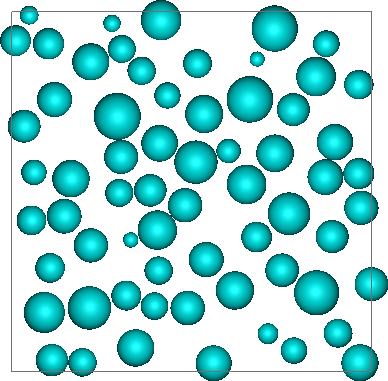}  
(b)\includegraphics[width=0.4\columnwidth]{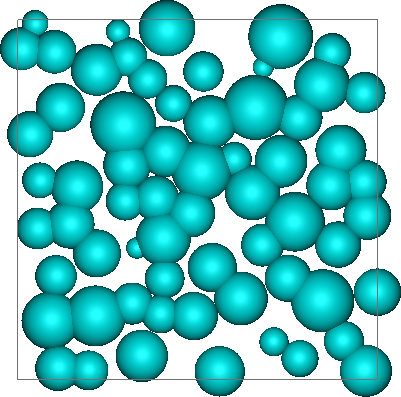} 
(c)\includegraphics[width=0.45\columnwidth]{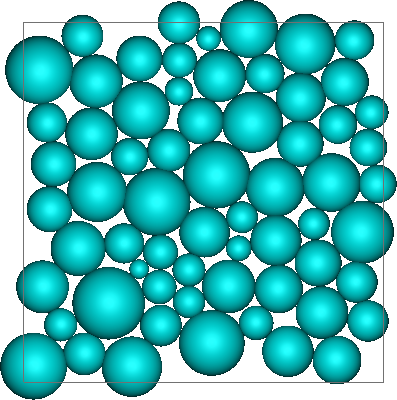}    
\caption{\label{fig2} Jammed state as generated by potential energy minimization. One start by picking a state (a) from an equilibrated hard-sphere fluid. Then, the soft shell interaction is turned-on  and the packing fraction is now such that there are many overlaps (b). The system relaxes then to the corresponding potential energy minimum, yielding the jammed state (c).  }
\end{figure}

The problem of generating model granular jammed states from an energy landscape can be divided into two parts. The first one consists in identifying a model whose energy minima mostly correspond to --- granular --- jammed states. The second one consists in providing a procedure for the landscape exploration that ultimately decides the frequency with which each jammed state is visited. For soft matter models, the potential energy surface is that of a system of hard core particles --- whose radii can go to zero in some models --- covered by a soft shell --- using hertzian \cite{Nagel03,Makse05, Xu11, Frenkel11}, elastic \cite{Makse12} or WCA \cite{PRL14} potentials ---. For our purposes, we recquire that the packing fraction of the whole system, when defined with respect to the total particle sizes ({\it i.e. hard core + soft shell}), is way above jamming. The ``\ above jamming ''\ condition ensures that all minima correspond to jammed states and not to a part of the configuration space that would be fluid-like. That way, the jammed basins tile certainly the underlying fluid space; which is a central feature of the method as we shall see below. The price to pay is then that the present analysis cannot straightforwardly be used to probe the whole energy landscape at the jamming transition. Study of this particular point below which most disordered hard particle assemblies are fluid \cite{Kamien07} goes beyond the scope of the present paper and is still the object of intensive studies \cite{Nagel03, Xu05, Majmudar07, Zhang2009, Makse10, Xu11, Jacquin11, Sollich12, Makse12, Charbonneau14}.\newline Now, regarding the exploration procedure, it is very common to choose a packing fraction for the hard core part of the system that is sufficiently low to avoid any glassy behaviour. In that case, if one picks, at random, configuration states from an equilibrated hard core fluid, then this is equivalent to exploring uniformly the potential energy surface associated to the soft shell interactions. This protocol is illustrated with typical simulation steps in Fig. \ref{fig2}. 

A consequence of this exploration protocol is that, upon a quench to the ``\ nearest ''\ minimum, a fluid state picked in this way is more likely to be captured by a minimum with a big basin of attraction rather than by a small one. In fact, the probability to end up in a particular basin of hyper-volume $\mathcal{V}$ is exactly $\mathcal{V}/\mathcal{V}_{fluid}$ with this aforementioned protocol; where $\mathcal{V}_{fluid}$ is the volume tiled by all the basins. Thus, in addition to being simple, this widely used protocol is also very convenient because we can infer straightforwardly what the jammed states statistics has to be (and it is not uniform).

\vspace{2mm}

Now, to prevent any confusion, a comment is here in order. The energy landscape description of soft material models of a granular medium discussed above (and in Fig.\ref{fig1}) looks very similar to the {\it free energy} landscape description of structural glasses. In fact, many studies have looked deep into the glassy regime with the use of the Replica Symmetry Breaking tools to probe the jamming transition for soft and hard particles ({\it e.g.} \cite{Zamponi06, Zamponi09, Jacquin11, Charbonneau12, Charbonneau14}). In particular, it has been found recently that when using the smallest resolution possible to define the basins of this free energy landscape for hard spheres, the actual hyper-volume of these basins was vanishing at jamming \cite{Charbonneau14}. This contrasts with the claim of the previous paragraph where the hyper-volume of the basins of attraction of jammed states have no reason to be zero (this is quite clear in Fig.\ref{fig1} too). These two observations are in fact non-contradictory for they do not talk about the same landscapes. On the one hand, the free energy landscape of structural glasses --- which is a Gibbs entropy landscape for hard spheres --- is an {\it emerging property} of the thermodynamic state of the system; that is, upon following a particular branch of a metastable fluid phase, some regions of the configuration space become more and more dynamically disconnected from one another as we go deeper into the glass phase. This gives rise to the successive appearance of more and more ``\ structural basins ''\ which characterize these effectively disconnected regions within which, in the glassy regime, particles are still free to move and hence a thermodynamic Gibbs entropy can be associated to them. It is then no surprise that, when approaching jamming from the glass, the ``\ typical ''\ entropy per basin vanishes because {\it by definition}, no particle has any freedom to jiggle at jamming and the Gibbs partition function has to be zero there. That is simply because, at jamming, the dimensionality of the space to explore decreases --- essentially from continuous to discrete \cite{Gao09} --- and as a consequence, the Gibbs measure in the continuous configuration space of hard spheres has to be exactly zero for any set in this new space of jammed states. On another hand, the potential energy surface discussed in Ref. \cite{PRL14} and in the present paper is a {\it feature} of a soft material model of a granular material and hence of the effective protocol used to generate model granular jammed states. The definition of basins of attraction --- and their hyper-volume --- then differs greatly in these two cases. The free energy landscape of glasses would define a basin as being the set of all fluid configurations {\it compatible} with a particular average structure (set that ought to be of cardinality one at jamming) whose hyper-volume is then often related to a typical {\it cage size} reminiscent of cell model methods used to compute the Helmholtz free energy of crystals. Essentially, the smaller the cage, the smaller the basin entropy and the harder it becomes to compress the system any further. Upon following a metastable branch of the fluid phase, this leads to a divergence of the thermodynamic pressure at a packing fraction much lower than that of close packing which is symptomatic of a vanishing typical cage size as discussed by Kamien and Liu \cite{Kamien07}. {\it A contrario}, the hyper-volume of a basin of attraction of a potential energy surface would be the set of all initial states which lead, upon a prescribed quenching dynamics, to the same jammed structure and has absolutely no reason to be of zero measure. For a glass, it could be the set of initial fluid states at an initial equilibrium packing fraction $\phi_i$ which, upon quenching to a higher final packing fraction $\phi_f$ beyond the glass transition, reach the same average structure. This is very different from the number of fluid states {\it at} $\phi_f$ which are compatible with some average structure. Now, in our case, looking at Fig. \ref{fig2}, the basin of attraction of the jammed structure $(c)$ is the set of all $(a)$-states leading to the structure $(c)$ upon following the steps decribed in the above paragraph and summarized in Fig. \ref{fig2}. This discussion on the difference between basin entropies in a glassy free energy landscape close to jamming and basin hyper-volumes of a potential energy surface echoes a more general discussion on other --- non-thermodynamic --- entropies that can be defined to characterize jammed states \cite{Daan13} and had to be made clear before moving on to the main point of the paper.

\section{Canonical ensemble for jammed states}

Ref. \cite{PRL14} aimed at estimating the {\it total} number of jammed states for a given packing fraction $\phi$ and number of particles $N$ (for a poly-disperse system). To do so, the authors imagine an {\it ideal} --- unknown --- protocol which would sample all of the jammed states uniformly and define the mean basin hyper-volume as being:
\begin{equation}
\langle \mathcal{V} \rangle \equiv \frac{1}{\Omega_{jammed}(N,\phi)}\sum_{\mathcal{B} \in \Omega_{jammed}} \: \mathcal{V}_{\mathcal{B}} \label{mean1} 
\end{equation}The idea is then to infer an unbiased distribution $p_u(F)$ for the {\it free energy} $F = -\ln (\mathcal{V}N!)$ \footnote{The additional $N!$ in the definition of the free energy of a basin of attraction ensures that the configurational entropy $\ln \Omega_{jammed}(N,\phi)$ (or other reasonable definitions of the configurational entropy) derived from \eqref{mean1} and \eqref{mean2} is extensive as discussed in length in Ref. \cite{PRL14}.} of a basin such that:
\begin{equation}
\langle \mathcal{V} \rangle = \int dF \: p_u(F) e^{-F} \label{mean2}
\end{equation}
The major problem is that, as specified before, $p_u(F)$ is unknown since the protocol described above to actually generate the jammed states is far from sampling them uniformly. At this stage, it is worth pointing out that the ideal unbiased algorithm that would sample all the jammed states uniformly has {\it a priori} nothing to do with the basin size distribution $p_u(F)$. Indeed, using an energy landscape is just a particular way --- among others --- to get jammed states. So, while we can always imagine jammed states as being minima of an energy landscape, we could have used another feature of jammed states to sample them uniformly, and that would not have changed their basin volume distribution. Let us now denote $p_b(F)$ the free energy distribution of the basins as sampled by the protocol with a bias proportional to the hyper-volume of the basins. It is clear that: 
\begin{equation}
p_b(F) \propto p_u(F) \mathcal{V}(F) = p_u(F) e^{-F}, \label{eq9}
\end{equation}and now something interesting happens. The expression in Eq.\eqref{mean2} can indeed be interpreted as a partition function similar to that of Eq.\eqref{eq0} associated to the biased weight in Eq.\eqref{eq9} with which each basin is sampled with the protocol in use. The interpretation goes then as follows: everything is as if we had ``\ microstates ''\ which are the jammed states and for each of them we can measure an ``\ energy ''\ which is nothing but the free energy associated to their corresponding basin of attraction ($F = -\ln (\mathcal{V}N!)$). In Eq. \eqref{mean2}, $p_u(F)$ plays exactly the same role as $\omega(E)$ in Eq. \eqref{eq0} corrected by a Boltzmann weight penalizing high energies in the case of Eq. \eqref{eq0} and small volumes in the case of Eq. \eqref{eq9}. In addition, $F$ is here strictly bounded from below because of the size of the box but has no obvious strict upper bound. A major difference with usual statistical mechanics though, is that both the biased {\it and} the unbiased distributions for $F$ have to be integrable and in general, we assume that most of their moments --- if not all --- have to be defined. A consequence of these general requirements is that the free energy density of state noted $\omega_{jammed}(F) \propto p_u(F)$ has to be a fast decaying function for both small and large values of $F$ (almost defining a compact support). 

\vspace{2mm}

Finally, for the analogy with a canonical ensemble to be complete, we can wonder what is the equivalent of $\beta$ in the case of jammed states (that we denote $\beta_G$). It turns out that $\beta_G$ is related to how sensitive the sampling protocol is to the size of the basins. With the protocol described above, the sampling selects proportionally to the hyper-volume of the basin and $\beta_G = 1$, but one could imagine other types of sampling for which the bias with respect to uniformity would be some power of the hyper-volume of those basins and this exponent would be the corresponding $\beta_G$.

\section{Towards a microcanonical ensemble}
One of the results of Ref.\cite{PRL14} is that the biased distribution $p_b(F)$ tends towards a gaussian distribution for large system sizes. It is also found that $\langle F \rangle_b \sim N f^* $ {\it i.e.}, the biased average of $F$ increases linearly with the system size \cite{PRL14}. In addition, it is found that $Var(F) \sim N c$ (where $c \sim \mathcal{O}(1)$) so that in the end, if we look at the biased distribution of the free energy per particle $f \equiv F/N$, it comes (for large system sizes):
\begin{equation}
p_b(f) = \sqrt{\frac{N}{2 \pi c}} \exp \left(- \frac{N (f-f^*)^2}{2c} \right) \label{eq10}
\end{equation}As we have seen before, such a function will tend towards a Dirac delta distribution and, in effect, will be uniformly sampling basins with a single free energy that is $Nf^*$. Now, the corresponding ensemble is one where the free energy is essentially $N f^*$ for every basin sampled. It is worth stressing that such a claim was already hypothesised by Wang et al. \cite{Makse12} to justify that Edwards' measure would be retrieved in the thermodynamic limit even for biased sampling protocols. Although we acknowledge that the shrinking of some basin-related probability distribution was indeed inspired, we disagree with the conclusion that the Edwards measure will always be retrieved in the thermodynamic limit and shall explain why below.

There are in fact two ways to interpret Eq. \eqref{eq10}: a) all possible jammed states are sampled uniformly because the landscape becomes such that all basins have the same hyper-volume corresponding to the free energy $Nf^*$ or b) only a subset of all possible states is sampled uniformly in the thermodynamic limit and that is the one comprising basins with a free energy $Nf^*$. The remaining part of the paper will deal with this issue.

\section{Finding the density of state upon unbiasing $p_b$}

Let us first remark that, if the biased distribution is genuinely gaussian (which is what we will assume here based on Ref. \cite{PRL14}), then so is the unbiased one. To see this, one simply needs to use Eq.\eqref{eq9} together with the following identity $-a(x-\overline{x})^2 + x = -a(x-(\overline{x}+(2a)^{-1}))^2 - \overline{x} - (4a)^{-1}$. As a matter of fact, since we multiply $p_b(F)$ by an exponential to retrieve $p_u(F)$, its mean will be shifted towards higher values of the free energy (an illustration of this is given {\it e.g.} in Ref. \cite{Daan13}). Now, the fact that the density of state of $F$ (or equivalently the probability distribution $p_u(F)$) tends to be gaussian for large system sizes was also found in \cite{Frenkel11} by sampling all the minima of small binary mixtures of elastic disks. This suggests that the gaussianity of $p_u(F)$ (or $\omega_{jammed}(F)$)  does not depend very much on the actual energy landscape as long as the interactions are kept short ranged and isotropic. 

In the end, for such energy landscapes, the free energy per particle distribution --- as measured by uniform sampling --- is a delta function centered around $f^* + \Delta$ where $\Delta$ accounts for this shift and is a positive number unimportant for the present discussion.

\vspace{2mm}

We already see here that if we believe in the interpretation a) above, that is, interpreting Eq. \eqref{eq10} as meaning that {\it all} states are sampled uniformly and have free energy $Nf^*$, then there is a contradiction with the fact that, when {\it all} the jammed states are sampled uniformly and we infer the corresponding unbiased distribution from $p_b$, then we know that there are overwhelmingly more states with free energy $N(f^* + \Delta)$ than with free energy $Nf^*$. Thus, interpretation a) cannot be right. The only remaining choice, common from a statistical mechanics point of view, is that to a canonical sampling with $\beta_G = 1$ corresponds an equivalent uniform sampling of a particular subset of all the jammed states that contains only jammed states with free energy $Nf^*$.


\vspace{2mm}

As said before, one can imagine protocols whose sensitivities to the basin size will go as $\mathcal{V}^{\beta_G}$. In such a case, the problem is equivalent to that of a canonical ensemble with an inverse ``\ temperature ''\ $\beta_G$ and a general partition function:
\begin{equation}
Q_G(N,\phi,\beta_G) \equiv  \int dF \: \omega_{jammed}(F)e^{-\beta_G F} \label{eq11}
\end{equation}
The crucial point to make here is that the gaussianity of $\omega_{jammed}(F)$ (or equivalently $p_u(F)$) for large $N$ is not protocol-dependent and is a feature of the soft matter model. As a consequence, any canonical weight with a temperature $\beta_G$, in virtue of the trivial extension of Eq. \eqref{eq9} to any $\beta_G$ value, will tend towards a gaussian for large system sizes and will be statistically equivalent to a uniform measure within a corresponding free energy slice in the thermodynamic limit.

\vspace{2mm}

In contrast to what was suggested in Ref. \cite{Makse12}, we stress that, in general, the limiting microcanonical distribution discussed above does not coincide with Edwards' ensemble as the latter requires that all possible states at a given $N$ and $\phi$ are uniformly sampled. In fact, Edwards' measure would correspond to the sum of $\omega(F)$ over all basin free energy slices {\it i.e.} over all the possible limiting microcanonical ensembles as derived in the present paper. Now, an interesting case is that of $\beta_G = 0$. This corresponds to an infinite temperature in our interpretation of the protocol and indeed in this case there is no bias and all states can be sampled uniformly; the Edwards measure is then retrieved as a very specific case of our interpretation. That being said, it is worth mentioning that an interesting discussion on how fast a quench is performed in an energy landscape and its effect on the statistical mechanics description of granular media has been proposed in \cite{Makse12}. It was indeed argued that slowly quenched protocols were compatible with Edwards' measure while, at least for small systems, fast quenches were not. This claim seems quite un-intuitive at first sight since we know that {\it e.g.} for hard spheres, a slow quench in packing fraction simply leads to an entropy-driven crystallization that ends up in a jammed close-packed FCC or HCP structure; this is not quite what one would call a uniform sampling of jammed states. This intuition is indeed confirmed by running standard Monte Carlo and Molecular Dynamics simulations but also LS simulations which, at sufficiently small compression rates, can mimic equilibrium behaviour \cite{Torquato2006, Torquato2011}. Now, the numerical quenching protocol of Ref. \cite{Makse12} (and also of Refs. \cite{Makse05, Makse08} with similar findings) actually differs from a usual LS algorithm in that it allows each particle to be subject to a background fluid-like dissipation that forces the system to eventually reach a mechanical equilibrium --- not necessarily jammed --- after each compression step \cite{Makse08}. Motion of the particles is then only triggered by particle overlaps upon further compression. Such a protocol is very similar in spirit to the one of Ref. \cite{Sollich12} cautiously designed to study the jamming of soft athermal systems. Hence, it is slightly misleading to interpret such a protocol as being related to a single potential energy surface --- that of the final packing fraction --- for the system explores many different energy landscapes along its compression trajectory. This phenomenology of ever-changing landscape upon reaching jamming bears some resemblance with what happens when trying to probe jamming from model spin glasses \cite{Krzakala10} or more recently hard sphere glasses \cite{Charbonneau14}, although, because the particles remain forever soft in the studies \cite{Makse08, Makse12}, a non-zero hyper-volume of the basins of attraction is expected (and indeed measured).

 In the present formalism, we claim that slowly reaching a minimum of an energy landscape (in the sense discussed in the previous paragraph) from an initial fluid configuration may in fact well be an artificial way of decreasing the value of $\beta_G$ towards zero but it is however a bit too hasty to conclude that the corresponding statistical measure is that of Edwards'. Instead, we propose that, as the protocol used is more and more able to effectively decrease $\beta_G$, one should observe more and more compatibility with Edwards' proposal and reach equality only in the limiting case $\beta_G = 0$. 


\vspace{2mm}

Finally, in case one were to be surprised that we can find an equivalent microcanonical measure --- centered at $N(f^*+\Delta)$ --- even when the sampling is unbiased and probing basins of all sizes {\it i.e.} when $\beta_G = 0$, it is worth recalling that the same goes for any thermodynamical system with a bounded energy spectrum in usual statistical thermodynamics. As a prototypical example, in a paramagnetic spin system coupled to an external magnetic field at inverse thermodynamic temperature $\beta$, one easily finds that the canonical ensemble at $\beta = 0$ is equivalent to a microcaninical ensemble where the energy $E = \langle E \rangle_{\beta = 0} = 0$. In this spin example, and in our granular case too, the unbiased sampling over all possible states is overwhelmingly dominated by the states with the most probable energy (or by basins with the most probable free energy in our granular case).

\section{Conclusion}
In this article, we have shown that it was possible to interpret the role of a widely used numerical protocol to generate jammed states as a particular ``\ granular''\ temperature value --- $\beta_G = 1$ --- of a much richer canonical ensemble. We have also shown that, assuming results from \cite{PRL14} can be extended to more general cases of energy landscape based protocols (cf. those used in Ref. \cite{Makse12}), this canonical ensemble is equivalent to a microcanonical ensemble in the thermodynamic limit where some jammed states are then sampled uniformly. We argued that Edwards' microcanonical ensemble was a very specific case of this richer picture where the protocol inverse temperature is $\beta_G = 0$, otherwise the other granular microcanonical measures $\beta_G > 0$ do not coincide with Edwards'.

\vspace{2mm}

These results shed light on how to take the role of the protocol seriously and yet be able to formulate the problem in a formalism identical to that of statistical thermodynamics. We hypothesise that the results discussed in this paper for energy landscape based protocols could extend to volume landscapes \cite{Ashwin12} and would give a new intuition about the statistical mechanics of tapping protocols for instance. It is worth noting that, even in the case of a volume landscape, the probability assigned to each jammed state would be directly linked to the hyper-volume of the corresponding basin of attraction. Only if it is possible to have a mapping from these hyper-volumes of basins of attraction to the corresponding packing fractions of the jammed states, can we try to infer from it some statistical predictions about the probability distribution of volume or packing fraction for a given protocol.\newline  Finally, the interpretation here introduced opens the window for novel granular simulations in which an energy landscape exploration could be coupled to a Metropolis criterion enforcing the value of $\beta_G$. This could give much more freedom in terms of protocols used to generate jammed states and interpret their physical properties.

\begin{acknowledgments}
{\small \it F. Paillusson acknowledges fruitful discussions with S. Martiniani, K.J. Schrenk, D. Frenkel and D. Asenjo. F.P. acknowledges the Direccion General de Investigacion (Spain) and DURSI for financial support under
projects FIS 2011-22603 and 2009SGR-634, respectively.}
 \end{acknowledgments}
\bibliographystyle{apsrev4-1}
\bibliography{biblio_granular}

\end{document}